# Average Sequence Dissimilarity under the multi-species coalescent with constant population size

Joseph Heled[1]

[1]Department of Computer Science, University of Auckland,Auckland, NZ

April 5, 2011

## 1 Average Sequence dissimilarity

The sequence dissimilarity between two aligned sequences is defined as the fraction of positions with non-equal characters. We will show how to analytically derive the average sequence dissimilarity (ASD) within and between species under a simplified multi-species coalescent setup. To generate the sequences, we first draw a random pure birth (Yule, 1925) species tree $S$ with $n$ taxa and birth rate $\lambda$. The population size for each branch is constant and equal to $N_e$. A gene tree $G$ compatible with $S$ is randomly drawn using the multi-species coalescent, and finally sequences for the tips of $G$ are simulated using the Jukes-Cantor (Jukes and Cantor, 1969) model with a mutation rate of $\mu$.

### ASD within species

Under those condition computing the ASD between two individuals within the same species is fairly straightforward. First, since the two lineages are always in a population of size $N_e$, regardless of species, the density of the time to coalescence is simply the exponential distribution with rate $1/N_e$ (Kingman, 1982)

$$f_{CW}(t) = \frac{1}{N_e} e^{-t/N_e}. \tag{1}$$

The evolutionary distance between two individuals whose common ancestor is $t$ time units in the past equals $2t\mu$, and so the dissimilarity under the JC model is

$$\text{JC}_\mu(t) = \frac{3}{4}\left(1 - e^{-\frac{4}{3}2\mu t}\right). \tag{2}$$

The ASD is given by integrating the product of density and value over time,

$$\text{ASD}_W(N_e, \mu, \lambda) = \int_0^\infty f_{CW}(t)\text{JC}_\mu(t)\text{d}t = \frac{6\mu N_e}{3 + 8\mu N_e}. \tag{3}$$

## ASD between species

The calculation of the ASD between species requires more work. While the dissimilarity as a function of time is given by equation (2) as before, the density of the time of coalescence depends on the Yule process. If the two populations diverged $x$ time units ago, the coalescence density of two lineages from those species would be $f_{CW}(t)$ shifted by $x$. Now, assuming the density of the divergence time of the two species is $f_Y(x)$, The density of two lineages from different species coalescing at time $t$ is obtained by integrating over all possible divergence times before the coalescence –

$$f_{CB}(t) = \int_0^t f_Y(x) f_{CW}(t-x)\text{d}x. \tag{4}$$

The ASD formula is similar to the within species case (Equation 3):

$$ASD_B(N_e, \mu, \lambda) = \int_0^\infty f_{CB}(t) JC_\mu(t)\text{d}t \tag{5}$$

We still need to figure out $f_Y(x)$. Since all species pairs are identical for our purposed, arbitrarily pick a pair $\{u, v\}$. First, partition all ranked species topologies with $n$ taxa into $n-1$ groups $\psi_1, \psi_2, \ldots, \psi_{n-1}$, where $\psi_i$ contains all topologies with $i$ ($1 \leq i \leq n-1$) surviving lineages at the time of the common ancestor of $u$ and $v$.

Let $c_i$ be the size pf $\psi_i$. To compute $c_i$, note that there are $\frac{(k-2)(k-3)}{2} + 2(k-2) = \frac{(k-2)(k+1)}{2}$ ways to coalesce $k$ lineages into $k-1$ without coalescing $u$ and $v$, giving,

$$c_i = \prod_{k=2}^{i} \frac{k(k-1)}{2} \prod_{k=i+2}^{n} \frac{(k-2)(k+1)}{2}$$
$$= \frac{(n-2)!(n+1)!}{2^{n-2}(i+1)(i+2)} \quad (6)$$
$$= \frac{n(n-1)\mathcal{R}_{n-1}}{(i+1)(i+2)}.$$

Where $\mathcal{R}_n$ is the total number of ranked trees with $n$ taxa. The density of a tree with a topology from $\psi_i$ and with height $h$ of the common ancestor of $\{u, v\}$ is (Joseph and Drummond (2011), Appendix C)

$$g_i(h) = \frac{1}{\mathcal{R}_n}(n-i)\binom{n}{i}\lambda e^{-\lambda(i+1)h}(1-e^{-\lambda h})^{n-i-1}. \quad (7)$$

Putting equations (6) and (7) together, we have the required density -

$$f_Y(x) = \sum_{i=1}^{n-1} c_i g_i(x). \quad (8)$$

After some very tedious expansions and simplifications, we can obtain a closed form formula

$$ASD_B(N_e, \mu, \lambda) = \frac{6\mu N_e}{8\mu N_e + 3} + \frac{3(n+1)}{4(n-1)(w-1)(w\lambda N_e + 1)} \left( (w+1)\binom{w+n}{n}^{-1} + w\frac{n-1}{n+1} - 1 \right), \quad (9)$$

where $w = \frac{8\mu}{3\lambda}$. Note that we use the extended binomial since $w$ is real.

Unlike the simple formula for within species, the ASD between species is not independent of $n$, and the number of terms increase with $n$ when the formula is expanded. For example, The ASD between species for $n = 3$ is

$$\frac{6\mu\left(32\mu^2 N_e + 60\lambda\mu N_e + 27\lambda^2 N_e + 12\mu + 18\lambda\right)}{(4\mu + 3\lambda)(8\mu + 9\lambda)(8\mu N_e + 3)} \quad (10)$$